\documentclass[prd,showpacs,preprintnumbers,amsmath,amssymb,11pt]{revtex4}

\usepackage{graphics}
\usepackage[utf8]{inputenc}
\usepackage{epsfig}
\usepackage{subfigure}
\usepackage{dcolumn}
\usepackage{bm}
\usepackage{color}

\begin{document}

\title{ Study on the radiative decays of $h_c$ via intermediate meson loops model}
\author{Qi Wu, Gang Li\footnote{gli@mail.qfnu.edu.cn}, Yawei Zhang}
\affiliation{College of Physics and Engineering, Qufu Normal University, Qufu 273165, China}

\begin{abstract}
Recently, the BESIII Collaboration reported two new decay processes $h_c(1P)\to \gamma \eta$ and $\gamma \eta^\prime$. Inspired by this measurement, we propose to study the radiative decays of $h_c$ via intermediate charmed meson loops in an
effective Lagrangian approach. With the acceptable cutoff parameter range, the calculated branching ratios of $h_c(1P)\to \gamma \eta$ and $\gamma \eta^\prime$ are orders of $10^{-4}\sim 10^{-3}$ and $10^{-3} \sim 10^{-2}$, respectively. The ratio $R_{h_c}=
\mathcal{B}( h_c\to \gamma \eta )/\mathcal{B}( h_c\to \gamma \eta^\prime )$ can reproduce the experimental measurements with the commonly acceptable $\alpha$ range. This ratio provide us some information on the $\eta-\eta^\prime$ mixing, which may be helpful for us to test SU(3)-flavor
symmetries in QCD.
\end{abstract}

\date{\today }
\pacs{13.25.GV, 13.75.Lb, 14.40.Pq}
\maketitle



\section{Introduction}
\label{sec:introduction}

The properties of charmonium states and their related theoretical ideas and methods which is based on theory of Quantum Chromodynamics(QCD) have already a lot of knowledge \cite{Voloshin:2007dx} since the first charmonium state $J/\psi$ was observed in 1974~\cite{Aubert:1974js,Augustin:1974xw}. All the charmonium states below $D\bar D$ threshold have been observed experimentally and can
be well described by potential models~\cite{Barnes:2005pb}. Among these states, the $P$-wave spin-singlet state $h_c(^1P_1)$ is the last charmonium state below the $D\bar D$ threshold that was confirmed experimentally. In $1992$, the E760 Collaboration at Fermi Lab first established this state in the $p\bar p$ annihilation. Since the quantum numbers of $h_c$ is $J^{PC}=1^{+-}$, it cannot be produced in $e^+e^-$ annihilation
directly. As a result, there are only a few decay modes of $h_c$ observed experimentally. The dominant decay mode of $h_c$ is $E1$ radiative transition and the branching ratio of $h_c\to \gamma \eta _{c}$ is about $(51 \pm 6)\%$~\cite{Rosner:2005ry,Dobbs:2008ec}. The hadronic decay $h_c\to 2(\pi^+\pi^-) \pi^0$ has a branching ratio $(2.2^{+0.8}_{-0.7})\%$~\cite{Adams:2009aa}, while the branching ratio of hadronic decay $h_c\to 3(\pi^+\pi^-) \pi^0$  only has an upper limit $<2.9\%$~\cite{Adams:2009aa}. Accordingly, there are not many theoretical studies of $h_c$. The $h_c$ production at hadron collider~\cite{Wang:2014vsa}, $e^+e^-$ annihilation~\cite{Wang:2012tz} and $B$ factory~\cite{Bodwin:1992qr,Beneke:1998ks,Jia:2012qx} are investigated. In Ref.~\cite{Li:2012rn}, authors studied the $O(\alpha_{s}v^{2})$ corrections to the decays of $h_{c}$ in non-relativistic QCD. In Ref.~\cite{Guo:2010zk}, Guo \textit{et al.} applied the NREFT to study the isospin
violation mechanisms of $\psi^\prime \to h_c\pi^0$. Liu and Zhao in Ref.~\cite{Liu:2010um} studied the helicity selection rule evading mechanism of the process $h_c$ decaying to baryon anti-baryon pairs with effective Lagrangian approach. Recently, Zhu and Dai in Ref.~\cite{Zhu:2016udl} studied the $\eta$ and $\eta^\prime$ production in the radiative $h_c$ decay with light-cone factorization approach.

Since the $h_c$ has negative $C$ parity, it very likely decays into a photon plus a pseudoscalar meson, such as $\eta_c$, $\eta$ and $\eta^\prime$. Very recently, based on the $4.48\times 10^8$ $\psi^\prime$ events collected with the BESIII detector operating at the BEPCII storage ring, the BESIII Collaboration firstly observed the radiative decay processes $h_c \to \gamma \eta$ and $\gamma \eta^\prime$  with a statistical significance of $4.0\sigma$ and $8.0\sigma$, respectively~\cite{Ablikim:2016uoc}. The measured branching fractions of $h_{c}\to \gamma \eta$ and $\gamma \eta^\prime$ are $(4.7\pm 1.5\pm 1.4)\times 10^{-4}$ and $(1.52\pm 0.27\pm 0.29)\times 10^{-3}$, respectively, where the first errors are statistical and the second are systematic uncertainties. These two decay modes may be useful for providing constraints to theoretical models in the charmonium
region. The ratio between them can also be used to study the $\eta-\eta^\prime$ mixing~\cite{Gilman:1987ax}, which is important to test SU(3)-flavor
symmetries in QCD.

In this work, we will investigate the radiative decays $h_{c}\to \gamma \eta(\gamma \eta^\prime)$ via intermediate meson loop(IML) model in an effective Lagrangian approach(ELA). IML transition is regarded as an important nonperturbative transition mechanisms which has a long history~\cite{Lipkin:1986bi,Lipkin:1986av,Lipkin:1988tg,Moxhay:1988ri} and recently are widely used to study the production and decays of ordinary and exotic states~\cite{Liu:2013vfa,Guo:2013zbw,Wang:2013hga,Cleven:2013sq,Chen:2011pv,Li:2012as,Li:2013yla,Voloshin:2013ez,Voloshin:2011qa,Bondar:2011ev,Chen:2011pu,Chen:2012yr,Chen:2013bha,Li:2015uwa,Li:2014gxa,Li:2014uia,Li:2013jma,Li:2013zcr,Li:2011ssa,Guo:2010ak,Wu:2016ypc,Wu:2016dws,Liu:2016xly,Li:2014pfa,Yuan-Jiang:2010cna,Zhao:2013jza,Li:2013xia,Wang:2012mf,Zhang:2009kr,Li:2007xr,Li:2007ky}. The paper is organized as follows: After the introduction in Sec.~\ref{sec:introduction}, we will present calculation of the radiative decays $h_{c}\to \gamma \eta(\gamma \eta^\prime)$ via the intermediate charmed meson loop and give some relevant formulas in Sec.~\ref{sec:Radiative decays}. In Sec.~\ref{sec:result}, the numerical results are presented. A brief summary will be given in Sec.~\ref{sec:summary}.

\section{The Radiative decays of $h_c$}
\label{sec:Radiative decays}

\begin{figure}[th]
\centering
\includegraphics[width=0.9\textwidth]{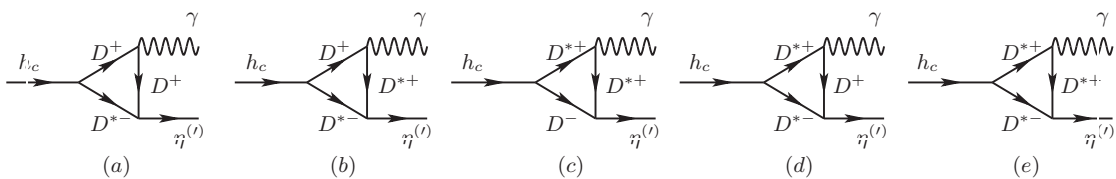}
\caption{The hadron-level diagrams for $h_c \to \gamma \eta$ and $\gamma \eta^\prime$
via charged intermediate charmed meson loops. Similar diagrams for neutral and strange intermediate charmed meson loops. }
\label{fig:feyn_hc}
\end{figure}
Generally speaking, we should include all the possible intermediate meson exchange loops in the calculation. In reality, the breakdown of the local quark-hadron duality allows us to pick up the leading contributions as a reasonable approximation~\cite{Lipkin:1986bi,Lipkin:1986av}. The coupling between $h_c$ and $D^{(*)} {\bar D}^{(*)}$ is an S-wave, so we consider the intermediate charmed meson exchange loops as the leading
contributions. At the hadronic level, as shown in Fig.~\ref{fig:feyn_hc}, the initial state $h_c$ dissolves into two charmed mesons which are off-shell and originated from the
coupled channel effects. Then these two virtual charmed mesons turn into final photon and $\eta(\eta^\prime)$ meson by exchanging the charmed meson.

In order to calculate the contributions from the charmed meson loops in Fig.~\ref{fig:feyn_hc}, we need the leading order effective
Lagrangians for the couplings. Based on the heavy quark symmetry~\cite{Colangelo:2003sa,Casalbuoni:1996pg}, the Lagrangian
for the P-wave charmonia at leading order is
\begin{equation}\label{eq:P-charmonium}
\mathcal{L}=ig_1Tr[P_{c{\bar c}}^\mu {\bar H}_{2i}\gamma _\mu {\bar H }_{1i}]+h.c. \, .
\end{equation}
where the spin multiplets for these four P-wave charmonium states are
expressed as
\begin{equation}
P_{c{\bar c}}^\mu=\frac{1 + \not v}{2} \left( \chi_{c2}^{\mu
\alpha }\gamma_\alpha +\frac{1}{{\sqrt 2}}\varepsilon^{\mu \nu \alpha
\beta }v_\alpha \gamma_\beta \chi _{c1\nu}+\frac{1}{{\sqrt 3}}(
\gamma^\mu -v^\mu ) \chi_{c0}+h_c^\mu \gamma_5 \right)
\frac{1-\not v}{2} \ ,
\end{equation}
with $v^\mu$ being the $4$-velocity of the multiplets.

The charmed and anti-charmed meson triplet read
\begin{eqnarray}
H_{1i} &=& \frac{1+\not v}{2} \left[ {\cal D}_{i}^{\ast \mu }\gamma _{\mu
}-{\cal D}_{i}\gamma _{5}\right] \, ,  \\
H_{2i} &=&\left[ {\bar{ {\cal D}}}_{i}^{\ast \mu }\gamma _{\mu }- {\bar {\cal D} }_{i}\gamma
_{5}\right] \frac{1-\not v}{2} \ , \\
{\bar H}_{1i,2i}&=&\gamma^0  H_{1i,2i}^\dag \gamma^0 \,
\end{eqnarray}
where ${\cal D}$ and ${\cal D}^{\ast }$ denote the pseudoscalar and vector charmed meson fields, respectively, i.e. ${\cal D}^{( \ast) }=\left( D^{0(\ast
) },D^{+( \ast ) },D_{s}^{+( \ast ) }\right)$. $v^\mu$ is the $4$-velocity of the charmed mesons. $\varepsilon_{\mu \nu
\alpha \beta }$ is the antisymmetric Levi-Civita tensor and $\varepsilon_{0123}=+1$.

Consequently, the relevant effective Lagrangian for $h_c$ reads
\begin{equation}
\mathcal{L}_{h_c {\cal D}^{( \ast ) }{\cal D}^{( \ast)
}}=g_{h_c {\cal D}^\ast {\cal D} }h_c^\mu \left( {\cal D} {\bar{\cal D}}_\mu^\ast + {\cal D}_\mu^\ast {\bar{ \cal D}} \right) +ig_{h_c {\cal D}^\ast {\cal D}^\ast }\varepsilon_{\mu \nu
\alpha \beta }\partial^{\mu }h_c^\nu {\cal D}^{\ast\alpha}{\bar{\cal D}}^{\ast\beta } \, ,
\end{equation}
where the coupling constants will be determined later.

The effective Lagrangian for light pseudoscalar meson coupled to charm mesons pair can be constructed based on the heavy quark limit and chiral
symmetry~\cite{Casalbuoni:1996pg,Colangelo:2003sa,Cheng:2004ru}
\begin{equation}
{\mathcal L}_{{\cal D}^{(\ast )}{\cal D}^{(\ast)} {\mathcal P}}=-ig_{{\cal D}^{\ast }{\cal D}
{\mathcal P}}\left( {\cal D}^i \partial^\mu {\mathcal P}_{ij} {\cal D}_\mu^{\ast
j\dagger }-{\cal D}_\mu^{\ast i}\partial^\mu {\mathcal P}_{ij} {\cal D}^{j \dag}\right) +\frac{1}{2}g_{{\cal D}^\ast D^\ast {\mathcal P}}\varepsilon _{\mu
\nu \alpha \beta }{\cal D}_i^{\ast \mu }\partial^\nu {\mathcal P}  {\overset{
\leftrightarrow }{\partial }}{\!^{\alpha }} {\cal D}_j^{\ast \beta\dag },
\end{equation}
where ${\mathcal P}$ is $3\times 3$ matrices for the pseudoscalar octet, i.e.,
\begin{eqnarray}
\mathcal{P} &=& \left(
\begin{array}{ccc}
\frac {\pi^0}{\sqrt{2}}+\frac {\eta\cos\alpha_P+\eta^\prime\sin\alpha_P} {%
\sqrt{2}} & \pi^+ & K^+ \cr \pi^- & -\frac {\pi^0} {\sqrt{2}}+\frac {
\eta\cos\alpha_P + \eta^\prime\sin\alpha_P} {\sqrt{2}} & K^0 \cr K^- & {\bar
K}^0 & - \eta\sin\alpha_P + \eta^\prime{\cos\alpha_P}
\end{array}
\right).
\end{eqnarray}

The physical states $\eta$ and $\eta^\prime$ are the linear combinations of $n{\bar n} = ({u\bar u} + {d\bar d})/\sqrt{2}$ and $s\bar{s}$ and they are taken to be the following form
\begin{eqnarray}
|\eta\rangle &=& \cos\alpha_P|n\bar n\rangle -\sin\alpha_P |s\bar s \rangle
\,  \notag \\
|\eta^\prime \rangle &=& \sin\alpha_P |n\bar n\rangle + \cos\alpha_P |s\bar
s \rangle \, , \label{Eq:eta_etap_mixing}
\end{eqnarray}
where $\alpha_P \simeq \theta_P + \arctan \sqrt{2}$. The empirical value for the
pseudoscalar mixing angle $\theta_P$ should be in the range $-24.6^\circ \sim -11.5^\circ$~\cite{Agashe:2014kda}. In this work, we will take $\theta_P
= -19.3^\circ$~\cite{Liu:2006dq} and $-14.4^\circ$~\cite{Ambrosino:2009sc}, respectively. The coupling constants will be determined in the next section.

In order to calculate these two radiative decay processes, the effective
Langrangian containing the interaction of photon are also needed.  If we implement the minimal substitution $\partial^{\mu }\rightarrow \partial^\mu +ieA^\mu$ for the  free scalar and massive vector fields, then we can obtain the relevant Lagrangians~\cite{Dong:2009uf,Mehen:2011tp},
\begin{eqnarray}
\mathcal{L}_{DD\gamma } &=&ieA_{\mu }D^{-}{\overset{\leftrightarrow }{%
\partial }}{\!^{\mu }}D^{+}+ieA_{\mu }D_{s}^{-}{\overset{\leftrightarrow }{%
\partial }}{\!^{\mu }}D_{s}^{+}\, , \\
\mathcal{L}_{D^{\ast }D^{\ast }\gamma } &=&ieA_{\mu }\left[ g^{\alpha \beta
}D_{\alpha }^{\ast -}{\overset{\leftrightarrow }{\partial }}{\!^{\mu }}%
D_{\beta }^{\ast +}+g^{\mu \beta }D_{\alpha }^{\ast -}\partial ^{\alpha
}D_{\beta }^{\ast +}-g^{\mu \alpha }\partial ^{\beta }D_{\alpha }^{\ast
-}D_{\beta }^{\ast +}\right] \notag \\
&& + ieA_{\mu }\left[ g^{\alpha \beta }D_{s\alpha }^{\ast -}{\overset{%
\leftrightarrow }{\partial }}{\!^{\mu }}D_{s\beta }^{\ast +}+g^{\mu \beta
}D_{s\alpha }^{\ast -}\partial ^{\alpha }D_{s\beta }^{\ast +}-g^{\mu \alpha
}\partial ^{\beta }D_{s\alpha }^{\ast -}D_{s\beta }^{\ast +}\right] \, ,
\end{eqnarray}
where $A\overleftrightarrow \partial_\mu B=A \partial_\mu B-(\partial_\mu A)B$, $F_{\mu \nu }=\partial_\mu A_\nu -\partial_\nu A_\mu $, and $M_{\mu \nu }=\partial_\mu M_\nu -\partial_\nu M_\mu $. Note
that the neutral interactions vanish. The interaction of $D^\ast D\gamma $ has the following form~\cite{Hu:2005gf,Amundson:1992yp}
\begin{eqnarray}
\mathcal{L}_{D^{\ast }D\gamma }= \frac{e}{4}\varepsilon^{\mu \nu \alpha \beta}F_{\mu \nu}[g_{D^{\ast +}D^{+}\gamma
}D_{\alpha \beta }^{\ast
+}D^+ + g_{D^{\ast 0}D^0\gamma }D_{\alpha \beta }^{\ast 0}D^0
+
g_{D_s^{\ast +}D_s^+\gamma }
D_{s\alpha \beta }^{\ast +}D_{s}^{+}]+H.c. \, ,
\end{eqnarray}

With the above Lagrangians, we can write out the explicit transition amplitudes of $h_c(p_1) \to [D^{(*)}(q_1) {\bar D}^{(*)}(q_3)] D^{(*)}(q_2) \to  \gamma (p_2) \eta^{(\prime)}(p_3)$ shown in Fig.~\ref{fig:feyn_hc},
\begin{eqnarray}
\mathcal{M}^{(a)} &=&\int \frac{d^4q_2}{( 2\pi )^4}
[ g_{h_cD^\ast D}\epsilon_{1\mu }] [ e\epsilon_{2 \theta }( q_2^\theta -q_1^\theta)] [
-g_{D^\ast D\eta }p_{3\rho }] \notag \\
&&\times \frac{i}{q_1^2-m_1^2}
\frac{i}{q_2^2-m_2^2}
\frac{i( -g^{\mu \rho }+q_3^\mu q_3^\rho /m_3^2) }{
q_3^2-m_3^2}\mathcal{F}( m_2,q_2^2) , \\
\mathcal{M}^{( b) } &=&\int \frac{d^4q_2}{( 2\pi )^4}
[ g_{h_cD^\ast D}\epsilon_{1\mu }] [
eg_{D^{\ast +}D^+\gamma }\varepsilon_{\theta \phi \kappa
\lambda }p_2^\theta \epsilon_2^\phi  q_2^\kappa ]
[ -g_{D^\ast D^\ast \eta }\varepsilon_{\rho \tau \sigma \xi
}p_3^\tau q_3^\sigma ]\notag \\
&&\times \frac{i}{q_1^2-m_1^2}\frac{i( -g^{\lambda\rho
}+q_2^{\lambda }q_2^\rho /m_2^2)}{q_2^2-m_2^2}\frac{i(-g^{\mu \xi
}+q_3^\mu q_3^\xi /m_3^2) }{q_3^2-m_3^2}\mathcal{F}
( m_2,q_2^2) , \\
\mathcal{M}^{( c) } &=&\int \frac{d^4q_2}{( 2\pi )^4}
[ g_{h_cD^\ast D}\epsilon_{1\mu }] [
e\epsilon_2^\theta [ g_{\kappa \lambda }( q_{2\theta
}-q_{1\theta }) +g_{\theta \lambda }q_{2\kappa
}-g_{\theta \kappa }q_{1\lambda }] ] [ g_{D^\ast D\eta }p_{3\rho }]  \notag \\
&&\times \frac{i(-g^{\mu \kappa }+q_1^\mu q_1^{\kappa }/m_1^2)}{
q_1^2-m_1^2}\frac{i(-g^{\lambda \rho }+q_2^\lambda q_2^\rho
/m_2^2)}{q_2^2-m_2^2}\frac{i}{q_3^2-m_3^2}\mathcal{F}
( m_2,q_2^2) , \\
\mathcal{M}^{( d) } &=&\int \frac{d^4q_2}{( 2\pi )^4}
\left[ g_{h_cD^\ast D^\ast }\varepsilon _{\mu \nu \alpha
\beta }p_1^\mu \epsilon_1^\nu  \right] [
eg_{D^{\ast +}D^+\gamma }\varepsilon_{\theta \phi \kappa
\lambda }p_2^\theta \epsilon_2^\phi  q_1^\kappa]
[-ig_{D^\ast D\eta }p_{3\rho}] \nonumber \\
 &&\times \frac{i(-g^{\beta \lambda }+q_1^\beta
q_1^\lambda/m_1^2)}{q_1^2-m_1^2}\frac{i}{q_2^2-m_2^2}\frac{
i\left( -g^{\alpha \rho }+q_3^\alpha q_3^\rho /m_3^{2}\right) }{
q_3^2-m_3^2}\mathcal{F}\left( m_2,q_2^2 \right) , \\
\mathcal{M}^{( e) } &=&\int \frac{d^4q_2}{( 2\pi )^4}
\left[ g_{h_cD^\ast D^\ast}\varepsilon_{\mu \nu \alpha
\beta }p_1^\mu \epsilon_1^\nu \right] \left[ e\epsilon_2^\theta
\left[ g_{\kappa \lambda }\left( q_{2\theta} -q_{1\theta}
\right) +g_{\theta \lambda }q_{2\kappa} -g_{\theta \kappa
}q_{1\lambda} \right] \right]  \notag \\
&&\times[-g_{D^\ast D^\ast \eta }\varepsilon_{\rho \tau \sigma \xi
}p_3^\tau q_3^\sigma ] \frac{i(-g^{\beta \kappa }+q_1^\beta
q_1^\kappa /m_1^2)}{q_1^2-m_1^2}\frac{i(-g^{\lambda \rho
}+q_2^{\lambda }q_2^\rho /m_2^2)}{q_2^2-m_2^2}  \notag \\
&&\frac{i(-g^{\alpha \xi }+q_3^\alpha q_3^\xi /m_3^2)}{
q_3^2-m_3^2}\mathcal{F}\left( m_2,q_2^2\right) \, ,
\end{eqnarray}
where $p_1$, $p_2$ and $p_3$ are the four momenta of the
initial state $h_c$, final state photon and $\eta (\eta^{\prime })$, respectively. $\varepsilon_1$ and $\varepsilon_2$ are the polarization vector of $h_c$ and photon, respectively. $q_1$, $q_3$ and $q_2$ are the four momenta of the
charmed meson connecting $h_c$ and photon, the charmed meson connecting $h_c$ and $\eta (\eta^{\prime })$, and the exchanged charmed meson, respectively.

In the triangle diagram of Fig.~\ref{fig:feyn_hc}, the exchanged charmed mesons are off shell. To compensate the offshell
effect and to regularize the divergence~\cite{Locher:1993cc,Li:1996cj,Li:1996yn}, we introduce a monopole form factor,
\begin{equation}
\mathcal{F}\left( m_2,q_2^{2}\right) =\frac{\Lambda^2-m_2^2}{
\Lambda^2-q_2^2} \, ,
\end{equation}
where $q_2$ and $m_2$ are the momentum and mass of the exchanged charmed meson, respectively.  The parameter $\Lambda \equiv
m_2+\alpha \Lambda_{QCD}$ and the QCD energy scale $\Lambda_{QCD} = 220 \mathrm{MeV}$. The determination of this dimensionless parameter $\alpha$ depends on specific process, which is usually of order $1$.

\section{Numerical Results}
\label{sec:result}
\begin{figure}[th]
\centering
\includegraphics[width=0.49\textwidth]{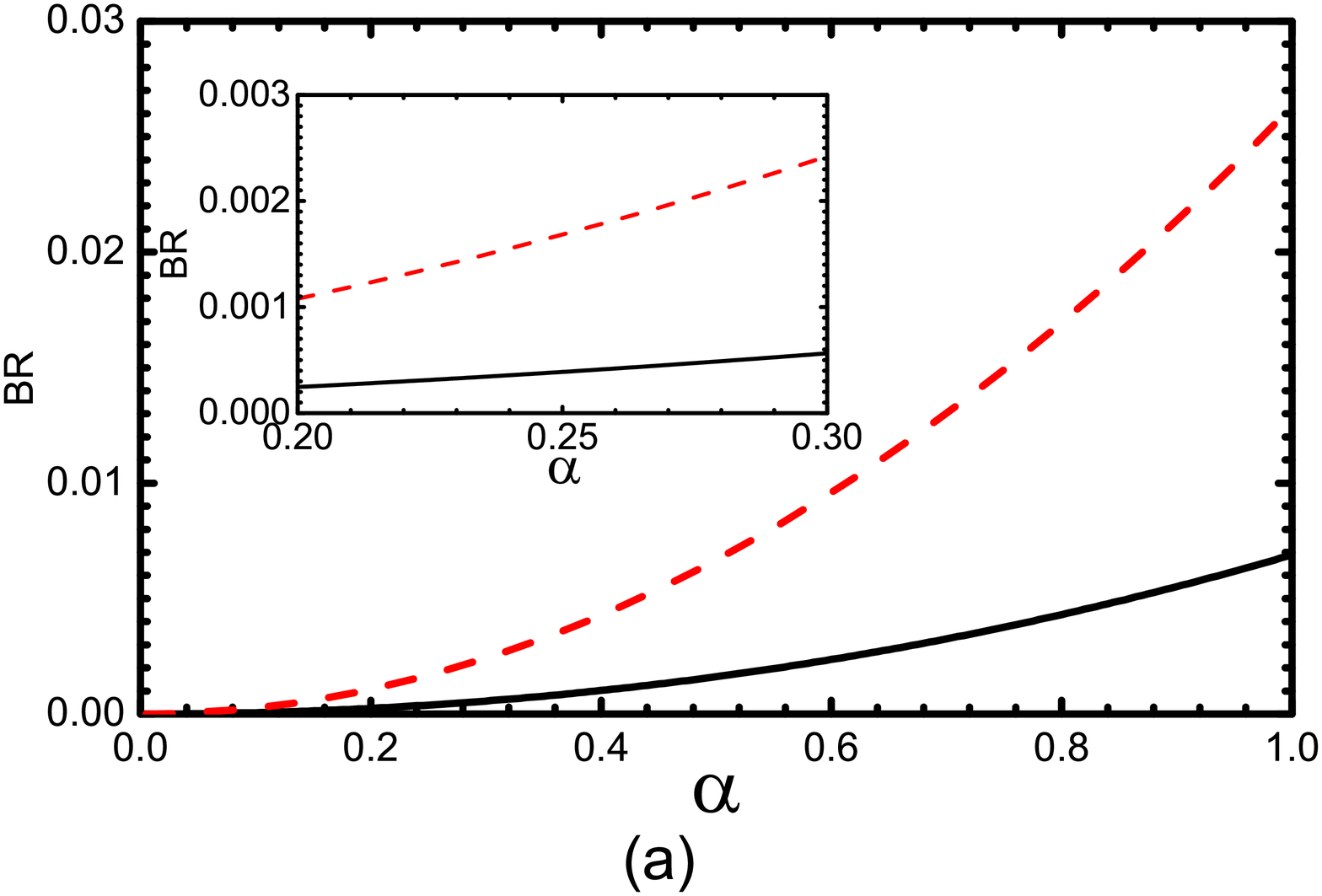}
\includegraphics[width=0.49\textwidth]{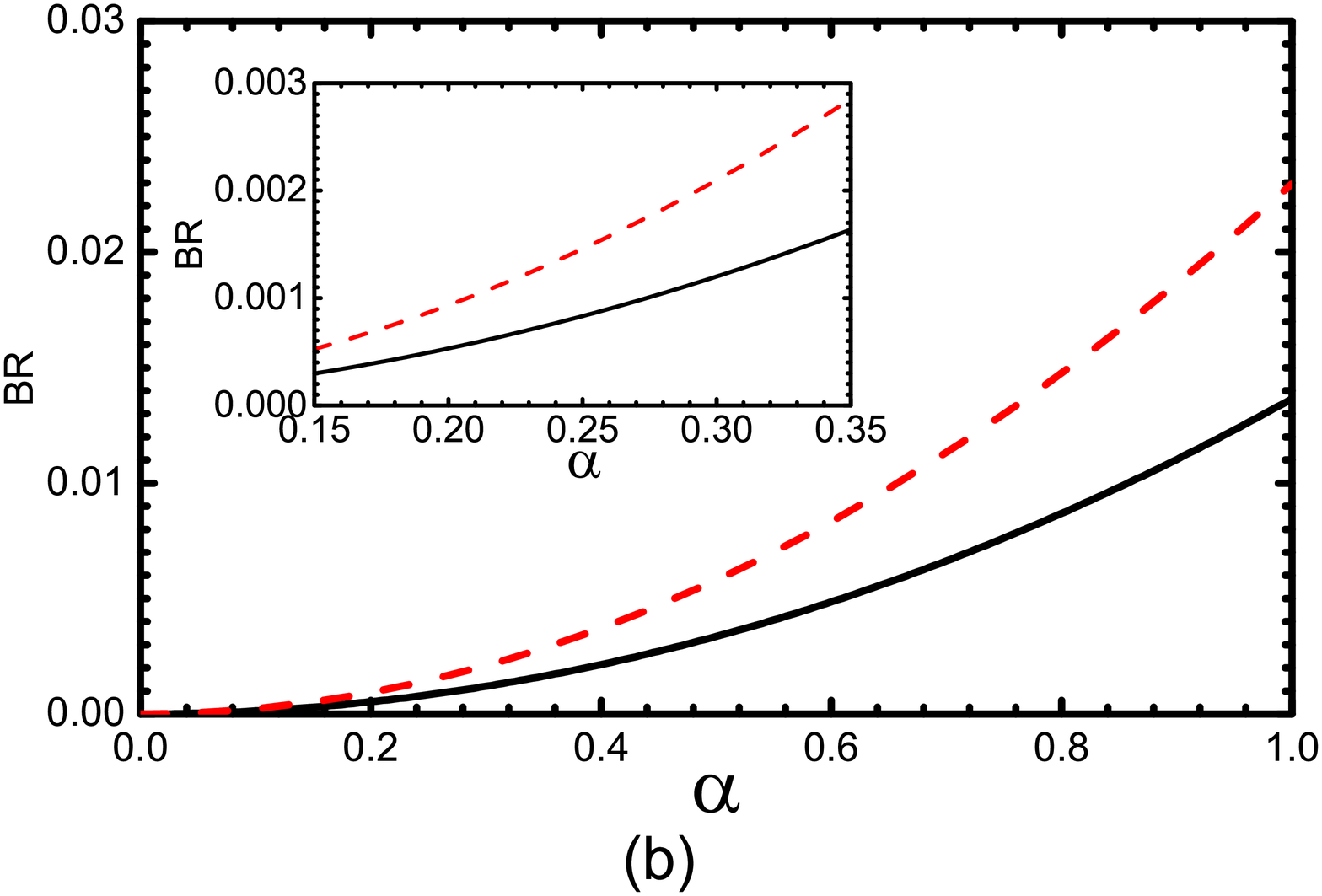}
\caption{ (color online). (a) The $\alpha $-dependence of the branching ratios of $h_c\to \gamma \eta$ (solid line) and $\gamma \eta^\prime$ (dashed line), respectively. The $\eta$-$\eta^\prime$ mixing angle $\theta_P=-19.3^\circ$ from Ref.~\cite{Liu:2006dq}. (b) The $\alpha $-dependence of the branching ratios of $h_c\to \gamma \eta$ (solid line) and $\gamma \eta^\prime$ (dashed line), respectively. The $\eta$-$\eta^\prime$ mixing angle $\theta_P=-14.4^\circ$ from Ref.~\cite{Ambrosino:2009sc}. }
\label{fig:br_alpha}
\end{figure}

\begin{figure}[th]
\centering
\includegraphics[width=0.49\textwidth]{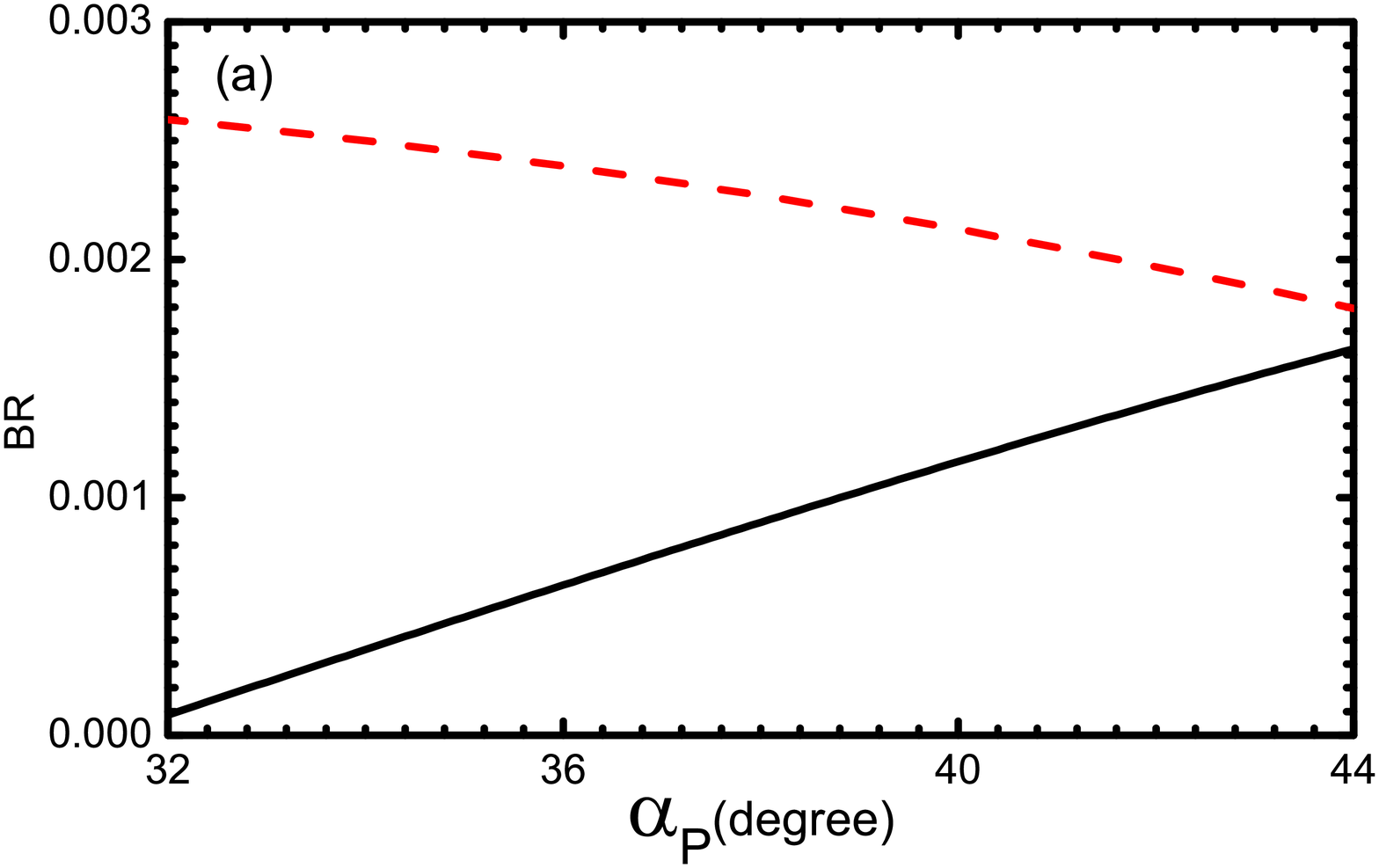}
\includegraphics[width=0.49\textwidth]{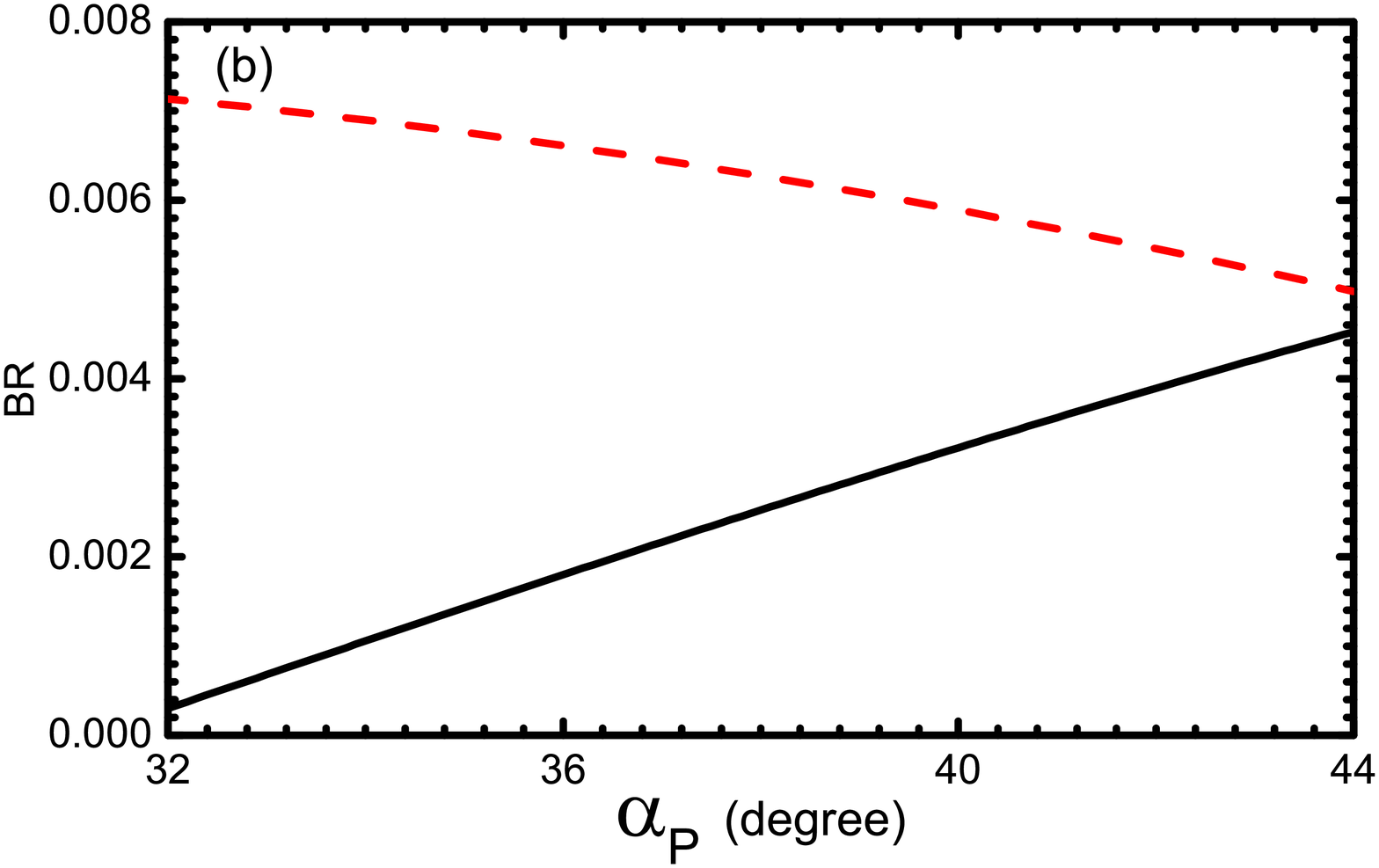}
\caption{ (color online). (a). The branching ratios of $h_c\to\gamma \eta$ (solid line) and $h_c\to\gamma \eta^{\prime}$ (dashed line) in terms of the $\eta$-$\eta^\prime$ mixing angle with $\alpha=0.3$.  (b). The branching ratios of $h_c\to\gamma \eta$ (solid line) and $h_c\to\gamma \eta^{\prime}$ (dashed line) in terms of the $\eta$-$\eta^\prime$ mixing angle with $\alpha=0.5$. }
\label{fig:br_mixing}
\end{figure}

\begin{figure}[th]
\centering
\includegraphics[width=0.6\textwidth]{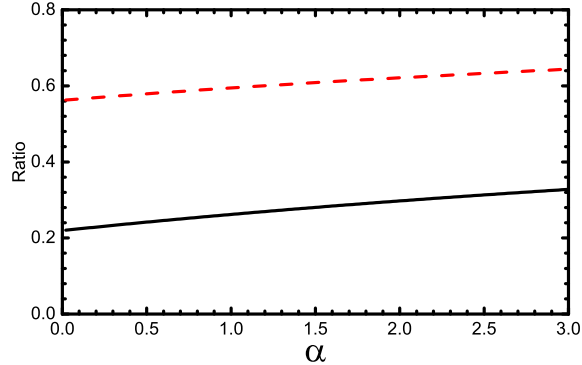}
\caption{(color online). The $\alpha $-dependence of the ratios $R_{h_c}$ with $\eta$-$\eta^\prime$ mixing angle $\theta_P=-19.3^\circ$ (solid line) and $\theta_P=-14.4^\circ$ (dashed line), respectively. }
\label{fig:ratio_alpha}
\end{figure}

\begin{figure}[th]
\centering
\includegraphics[width=0.6\textwidth]{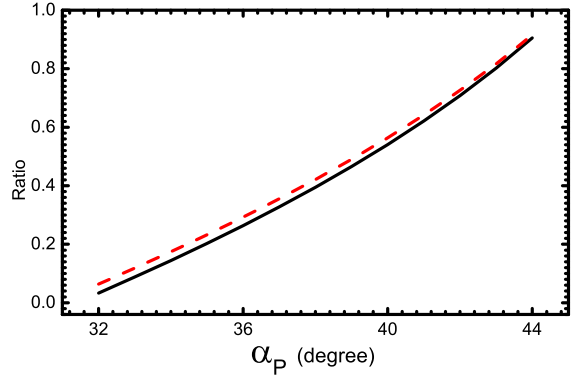}
\caption{The Ratios $R_{h_c}$ in terms of the $\eta$-$\eta^\prime$ mixing angle with $\alpha=0.3$ (solid line) and $\alpha=1.0$ (dashed line).}
\label{fig:ratio_mixing_angle}
\end{figure}

The coupling constants $g_{h_c{\cal D}^*{\cal D}}$ and $g_{h_c{\cal D}^*{\cal D}^*}$ are determined as
\begin{equation}
g_{h_c{\cal D}^*{\cal D}}=-2g_{1}\sqrt{m_{h_c}m_{\cal D} m_{{\cal D}^*}} , \quad g_{h_c{\cal D}^{\ast}{\cal D}^{\ast
}}=2g_{1}\frac {m_{{\cal D}^*}}{\sqrt{m_{h_c}}},
\end{equation}
with $g_{1}=-\sqrt{m_{\chi_{c0}}/3}/f_{\chi _{c0}}$, where $m_{\chi _{c0}}$
and $f_{\chi_{c0}}=510\pm 40$ $\mathrm{MeV}$ are the mass and decay
constant of $\chi_{c0}$, respectively~\cite{Colangelo:2003sa}.

In the heavy quark and chiral limits, the charmed meson
couplings to pseudoscalar mesons have the following~\cite{Cheng:2004ru},
\begin{equation}
g_{{\cal D}^\ast {\cal D}\mathcal{P}}=\frac{2g}{f_\pi }\sqrt{m_{\cal D} m_{{\cal D}^\ast }}\, , \quad g_{{\cal D}^\ast {\cal D}^\ast \mathcal{P}}=\frac{g_{D^\ast D\mathcal{P}}}{
\sqrt{m_{\cal D}m_{{\cal D}^\ast }}}\, ,
\end{equation}
where $g=0.59,$ $f_\pi =132$ $\mathrm{MeV}$ are adopted.

With the help of the measured experimental total width of $D^{*+}$ and the branching ratio of $D^{\ast +}\rightarrow D^{+}\gamma$~\cite{Agashe:2014kda}, we determine the coupling constant $g_{D^{*+}D^+\gamma}=0.5 \mathrm{GeV}^{-1}$. Since the $D^{*0}$ and $D_s^{*\pm}$ total widths are kept unknown,
we adopt the following values $g_{D^{* 0}D^0\gamma } \simeq 2.0 \mathrm{GeV}^{-1}$~\cite{Dong:2008gb} and$g_{D_{s}^{\ast }D_{s}\gamma }=-0.3\pm 0.1 \mathrm{GeV}^{-1}$\cite{Zhu:1996qy}.

In Fig.~\ref{fig:br_alpha} (a), we plot the $\alpha$ dependence of the branching ratios
of $h_c\to \gamma \eta$ (solid line) and $h_c\to \gamma \eta^\prime$ (dashed line) with $\theta_p=-19.3^\circ$, respectively. We also zoom in detail of the figure with a narrower range $\alpha=0.2\sim 0.3$ in order to show the best fit of $\alpha$ parameter. As shown in this figure, there is no cusp structure in the curve which is because the mass of $h_c$ lies
below the intermediate $D{\bar D}^*$ threshold. The $\alpha $ dependence of the branching ratios are not drastically sensitive with commonly accepted $\alpha$ range.  For the process $h_c\to \gamma \eta$, our calculated branching ratios can reproduce the experimental
data~\cite{Ablikim:2016uoc} at $\alpha =0.27\pm 0.06$. For $h_c\to \gamma \eta^\prime$, the results are consistent with the experimental measurements with $\alpha =0.24\pm 0.03$. At the same cutoff parameter $\alpha$, the calculated branching ratios of $h_c\to \gamma \eta^\prime$ are about 1 orders of magnitude larger than that of $h_c\to \gamma \eta$, which is mainly attribute to the $\eta$-$\eta^\prime$ mixing shown in Eq.~(\ref{Eq:eta_etap_mixing}). In Fig.~\ref{fig:br_alpha} (b), with $\theta_P=-14.4^\circ$, we plot the $\alpha$ dependence of the branching ratios
of $h_c\to \gamma \eta$ (solid line) and $h_c\to \gamma \eta^\prime$ (dashed line), respectively. We also zoom in detail of the figure with a narrower range $\alpha=0.15\sim 0.35$ in order to show the best fit of $\alpha$ parameter. The behavior is similar to that of Fig. 2(a). With $\theta_P=-14.4^\circ$, the branching ratios of $h_c\to \gamma \eta$ and $\gamma \eta^\prime$ can reproduce the experimental data with $\alpha=0.188_{-0.048}^{+0.038}$ and $0.26_{-0.03}^{+0.02}$, respectively. The errors for $\alpha$ are asymmetric. This asymmetry comes from a fact that the $\alpha$ dependence of the ${\cal B} (h_c \to \gamma \eta(\eta^\prime))$ is nonlinear.

In order to illustrate the impact of the mixing angle, in Fig.~\ref{fig:br_mixing}(a) and (b), we present the branching ratios in terms of the $\eta$-$\eta^\prime$ mixing angle with $\alpha=0.3$ (solid line) and $0.5$ (dashed line), respectively. In the case $\alpha=0.3$, when the mixing angle $\alpha_P$ increase, the branching ratios of $h_c \to \gamma \eta$ increase while the branching ratios of $h_c \to \gamma \eta^\prime$ decrease. This behaviour suggests how the mixing angle influences our calculated results to some extent. A similar behavior appears in the case $\alpha=0.5$.

As is well known, the $\eta$-$\eta^\prime$ mixing is a long-standing question in the literature. This mixing angle plays an important role in physical processes involving the $\eta$ and $\eta^\prime$ mesons. In Ref.~\cite{Ablikim:2016uoc}, the BESIII Collaboration measured the branching fraction ratio $R_{h_c}=
[\mathcal{B}( h_c\to \gamma \eta )/\mathcal{B}( h_c\to \gamma \eta^\prime)] = [30.7\pm 11.3(stat)\pm 8.7(sys)]\%$. This ratio $R_{h_c}$
can be used to study the $\eta -\eta^\prime$ mixing~\cite{Gilman:1987ax}, which is important to test SU(3)-flavor symmetries in QCD. In Fig.~\ref{fig:ratio_alpha}, we plot the $\alpha$ dependence of the ratio $R_{h_c}$ with $\theta_P=-19.3^\circ$ (solid line) and $-14.4^\circ$ (dashed line), respectively. As shown from this figure, the calculated ratio $R_{h_c}$ can reproduce the experimental measurements at the commonly acceptable $\alpha$ range for $\theta_P=-19.3^\circ$.  With $\theta_P=-14.4^\circ$, the calculated ratio $R_{h_c}$ is slightly larger than the experimental value. Furthermore, this ratio is less sensitive to the cutoff parameter $\alpha$, which is because the involved loop are same. When we take the ratio, the coupling vertices are cancelled out, so the ratio reflects the open threshold effects through the intermediate charmed meson loops and the mixing angle between $\eta$ and $\eta^\prime$ to some extent. In Fig.~\ref{fig:ratio_mixing_angle}, we plot the  $\eta$-$\eta^\prime$ mixing angle dependence of the ratios $R_{h_c}$ at $\alpha=0.3$ (solid line ) and $1.0$ (dashed line), respectively. This ratio changes very small when increasing the cutoff parameter $\alpha$, as a result, it can be used to probe the $\eta -\eta^\prime$ mixing. In our study, at $\alpha=0.3$, our results are consistent with the experimental measurements in the range $\alpha_P=(36.7^{ +2.1}_{-2.3})^\circ$, which corresponds to $\theta_P=(-18.0^{+2.3}_{-2.1})^\circ$. In the case $\alpha=1.0$, we can reproduce the experimental data in the range $\alpha_P=(36.2^{+2.2}_{-2.4})^\circ$, which corresponds to $\theta_P=(-18.5^{+2.4}_{-2.2})^\circ$. So our calculations can give a strong constrain on the $\eta$-$\eta^\prime$ mixing angle and we expect more precise measurements on this ratio, which may help us constrain this mixing angle.

The $\eta$-$\eta^\prime$ mixing angle can neither be calculated from the first principles in QCD nor measured from experiments directly. There are a lot of studies on this subject using different
methods~\cite{Leutwyler:1996sa,Gerard:2004gx,Schechter:1992iz,Feldmann:1998vh,Escribano:2008wi} and different processes, including various decay processes involving the light
pseudoscalar mesons. For example, in Ref.~\cite{Ambrosino:2009sc}, the KLOE collaboration updated the $\eta$-$\eta^\prime$ mixing angle value
by fitting their measurement $R_\phi = BR(\phi \to \gamma \eta)/BR(\phi \to \gamma \eta^\prime)$ together with several other decay channels. From the fit they extract the $\eta$-$\eta^\prime$ mixing angle $\theta_P=(-14.4 \pm 0.6)^\circ$. In Ref.~\cite{Guo:2015xva}, authors studied the $\eta$-$\eta^\prime$
mixing up to next-to-next-to-leading-order in U(3) chiral perturbation theory in the light of recent lattice simulations and phenomenological inputs. Within the framework of the effective Lagrangian approach, authors perform a thorough analysis of the $J/\psi \to VP$, $J/\psi \to \gamma P$ together with a few other processes to investigate this mixing problem~\cite{Chen:2014yta}. In the future, more decay processes involving the light
pseudoscalar mesons and more precise experimental measurements may will provide a unique method to study the $\eta$-$\eta^\prime$ mixing effects deeply.

\section{Summary}
\label{sec:summary}

In this work, we investigate the radiative decay processes $h_c \to
\gamma\eta$ and $\gamma \eta^\prime$ via intermediate meson loop model in an effective
Lagrangian approach. Our results show that the
obtained branching ratios are not drastically sensitive to
the cutoff parameter $\alpha$ to some extent. The calculated branching ratios of $h_c\to \gamma \eta$ are typically at the order of $10^{-4}\sim 10^{-3}$, while for $h_c\to \gamma \eta^\prime$, the branching
ratios are of order of $10^{-3}\sim 10^{-2}$ in the same cutoff range. The study of these two decay channels, especially their ratio $R_{h_c}$ can provide us some information on the $\eta$-$\eta^\prime$ mixing, which may be helpful for us to test SU(3)-flavor
symmetries in QCD.  The BESIII detector will collect $3\times 10^9$ $\psi^\prime$ events~\cite{Asner:2008nq}, which will provide a unique method to study the $\eta$-$\eta^\prime$ mixing effects deeply.

\section*{Acknowledgements}
The authors are very grateful to Qiang Zhao for useful discussions. This work is supported in part by the National Natural
Science Foundation of China (Grants no.11675091).

\end{document}